\documentstyle[preprint,aps]{revtex}

\tightenlines

\begin{document}
\draft

\preprint{UT-Komaba/96-23 , KEK-TH-498}

\title{Manifestly T-Duality  Symmetric Matrix Models}

\author{
Tsunehide Kuroki$^1$, \,  
Yuji Okawa$^1$
\footnote{JSPS Research Fellow}, \,
Fumihiko Sugino$^2$, \, 
Tamiaki Yoneya$^1$
 }

\address{$^1$ Institute of Physics, University of Tokyo, \\
Komaba, Meguro-ku, Tokyo 153, Japan \\
{\tt kuroki, okawa, tam@hep1.c.u-tokyo.ac.jp}\\
$^2$KEK Theory Group, Tsukuba, Ibaraki 305, Japan\\ 
{\tt sugino@theory.kek.jp}
}

\maketitle

\begin{abstract}

\end{abstract}
We present a new class of matrix models which are 
manifestly symmetric under the T-duality transformation 
of the target space. The models may serve as a 
nonperturbative regularization for the T-duality symmetry in  
continuum string theory. In particular, it now becomes possible to 
extract winding modes explicitly in terms of extended matrix variables. 

\newpage
\section{Introduction}
Matrix models can be regarded as a nonperturbative 
regularization of string theories. 
Recently, it became clear that various 
duality symmetries play vital roles in understanding the 
nonperturbative properties of string theories.  
Unfortunately, however, it does not seem that 
the standard matrix models are particularly suited for 
studying the duality symmetries.  Even  the T-duality  which 
is valid to each order of genus expansion in continuum string 
theories is difficult to 
implement manifestly in matrix models, at least in their 
present interpretation. For example, 
as discussed in our previous paper \cite{akosy}, 
the Ising model described by a standard 
two-matrix model 
does not preserve T-duality symmetry, whose validity is naively 
expected from the Kramers-Wannier duality on a square lattice, 
once we take into account the higher genus effects.  
The reason for this failure is that there is 
no symmetry between global winding and momentum modes in such  
models. 
Although it is not difficult to define partition functions as sums over 
random surfaces preserving T-duality 
\footnote{
See, e.g., \cite{gross-kleb} in which 
the case of $c=1$ has been discussed. 
}, it is quite nontrivial 
to write down corresponding matrix models preserving 
 T-duality symmetry manifestly.   
Since at present  matrix-model approaches 
 seem to be the only known tractable way   
towards {\it dynamical} formulation of 
nonperturbative string theories, 
it is  worthwhile to develop methods of 
treating duality symmetries exactly  
using matrix models from various 
possible standpoints.  
This is particularly so in view of a recent trend 
concerning new possibilities of using matrix models 
in seeking for fundamental theories of strings including 
D-branes.    

The purpose of the present paper is to present 
a new class of extended matrix models 
which have exact and manifest T-duality symmetry 
under the usual random-surface interpretation of 
matrix models. Although the 
models in general are not exactly soluble using presently available 
techniques, we believe that the existence of such models is
interesting by itself and 
will be useful for future investigations.  
To maintain the T-duality symmetry exactly, we need two 
key ingredients 
that have to be combined in a suitable way. 
One is that the global, as well as local, 
degrees of freedom corresponding to the 
winding and momentum modes must appear symmetrically for arbitrary 
topology of surfaces. 
The other is that the random triangulation 
of surfaces must be invariant under the duality transformation. 
For the first property, we adopt a variant of $Z_K$ spin systems  
($K=$ an integer) 
as the target space, generalizing the discussion given  
in the Appendix A of our previous paper \cite{akosy}.  
For the second, we can use the result of 
reference \cite{siegel} which is, to the 
best of our knowledge, the only known matrix 
model with exact self-duality.  In this 
reference,  the dual transformation for the 
random triangulation (namely, the measure of pure 2D gravity)
 was discussed, but the first problem was 
not addressed.  In fact, a naive generalization of the 
method used for  the one-matrix model 
  does not work for systems coupled with matter. 
We will clarify how to combine self-dual matter systems with 
the method of \cite{siegel}.  

The plan of the paper is as follows.  In the next section, we 
briefly discuss a class of $Z_K$ spin systems on a fixed surface, 
which we call $Z_{P, Q}$ 
models ($K=PQ$) and have manifest symmetry between 
``momentum" and ``winding" modes analogously with the 
toroidal compactification of continuum string theory. 
In section III, we present our extended matrix models by coupling 
the $Z_{P,Q}$ models to 2D gravity with the self-dual measure, 
and exhibit their exact T-duality symmetry. 
In section IV, some simple cases will be discussed for the 
purpose of concrete illustration. We will then discuss some 
relevant issues including the $c=1$ limit.  

\section{$Z_{P,Q}$ model on a fixed surface}

As is well known, $Z_K$ spin systems in two dimensions 
\cite{alcarazcardy}  
are self-dual under the Kramers-Wannier dual transformation.  
There is, however, a subtlety when we do not neglect 
global degrees of freedom. To make the system 
completely self-dual, we have to impose appropriate 
constraints in order to suppress 
the local vortex excitations which tend to 
violate the duality symmetry. In the case of the lattice 
$c=1$ matter system, this was discussed 
in \cite{gross-kleb} and the Appendix A of \cite{akosy}. 
Here we present its generalization to $Z_K$ systems. 
For clarity of notation, we assume the surface to be 
a fixed square lattice  
and denote the spin sites by $x$ and links  
by $x,\mu$ where  $\mu$ is the direction 
of the link from a site $x$. The difference operation 
between the nearest neighbor 
sites $x$ and $x+\hat{\mu}$ where $\hat{\mu}$ 
is a unit lattice vector along the  $\mu$ direction is denoted 
by $\Delta_{\mu}$. 

For a given $K$, we introduce two integers $P, \, Q$ 
satisfying $K=PQ$. 
Spin variables $n_x=0, 1, \ldots , P-1 $ (mod $P$) 
 are then assumed to live in the space $Z_P$. 
In addition to the spin variables, we introduce link variables 
$m_{x,\mu}=0, 1, \ldots, Q-1$ (mod $Q$)  living in the space $Z_Q$. 
Then the partition function of our $Z_{P,Q}$ model is 
defined as 
\begin{equation}
Z=\bigl(
\prod_{x, \mu}\, \sum_{n_x,  m_{x,\mu}}\bigr) 
\prod_x \delta^{(Q)}
(\triangle_{\mu} m_{x, \nu}-\triangle_{\nu}m_{x, \mu})
\prod_{x, \mu}
B(\triangle_{\mu}n_x - P m_{x, \mu}).
\label{partitionfunction}
\end{equation}
The Boltzmann factor $B(x)$ is assumed to be periodic 
under the translation $x\rightarrow x+ K$,  
and the $\delta$-function constraint is understood  modulo $Q$. 
Note that after the summation over the link variables, the Boltzmann 
factor can be regarded as a function on $Z_P$ 
with respect to the spin variables.  
Obviously, if the topology of the fixed surface is sphere, 
the constraint 
leads to 
\begin{equation}
m_{x, \mu} = \Delta_\mu m_x , \, \, m_x \in Z_Q. 
\label{spheresol}
\end{equation}
Therefore by redefining the spin variable by 
$n_x \rightarrow n'_x =n_x -P m_x \in Z_{PQ}=Z_K$, 
the system is reduced to an ordinary $Z_K$ spin system.  
If the fixed surface is not the sphere, 
we have instead of  (\ref{spheresol}) 
\begin{equation}
m_{x, \mu} = \Delta_{\mu}  m_x + \overline{m}_{x,\mu} , 
\, \, (m_x, \overline{m}_{x,\mu}) \in Z_Q ,
\end{equation}
where $\overline{m}_{x, \mu}$ is a global vector field 
which cannot be reduced to the difference of the site variables 
and is associated with a  
nontrivial homology cycle of the surface. The degree of  freedom 
represented by $\overline{m}_{x,\mu}$ is the analogue of the 
winding modes in the toroidal compactification 
of continuum string theory.  

Now let us 
perform the dual transformation on the partition function 
(\ref{partitionfunction}).  We introduce two auxiliary fields 
$\tilde{n}_y \in Z_Q$ and $\psi_{y, \mu}\in Z_K$ on 
dual sites $y$ and dual links, respectively, 
and rewrite (\ref{partitionfunction}), 
apart from a numerical proportional factor, as 
\begin{equation}
Z=\bigl(
\prod_{x, \mu}\,  \sum_{n_x, m_{x,\mu}}
\prod_{y, \mu} \sum_{\tilde{n}_y, \psi_{y,\mu}}\bigr)
\prod_x {\rm e}^{i[{2\pi \over Q}\tilde{n}\epsilon_{\mu\nu}
\Delta_{\mu}m_{\nu} + {2\pi \over K}\epsilon_{\mu\nu}\psi_{\mu}
(\triangle_{\nu}n - P m_{\nu})]}
\prod_{y, \mu}
\tilde{B}(\psi_{y,\mu}).
\end{equation}
Here we suppressed the 
subscripts for the sites on the exponential 
to avoid unnecessary notational complexity.  
The Boltzmann factor $\tilde{B}$ is the $Z_K$ Fourier transform 
of the original Boltzmann factor:
\begin{equation}
B(a) ={1\over \sum_{b=0}^{K-1}\tilde{B}(b)}
\sum_{b=0}^{K-1}{\rm e}^{i{2\pi \over K}ab}\tilde{B}(b). 
\label{fouriertransboltzmann}
\end{equation}
Reality of the Boltzmann  factor leads $\overline{\tilde{B}(b)}
=\tilde{B}(K-b)$. We assume the normalization condition 
$B(0)=\tilde{B}(0)=1$.   
Then solving the constraint coming from the summation over $m_{\mu}$, 
the general solution for $\psi_{\mu}$ is 
\begin{equation}
\psi_{\mu} = \Delta_{\mu}\tilde{n}- Q\tilde{m}_{\mu}, 
\, \, \tilde{m}_{\mu}\in Z_P. 
\end{equation}
After substituting this result and taking the summation over 
the original spin variables $n$, we have 
the constraint, 
\begin{equation}
\epsilon_{\mu\nu}\Delta_{\mu}\tilde{m}_{\nu}=0  
\,  \, \, ({\rm mod} \, P). 
\end{equation}
Thus the partition function after dual transformation takes, 
apart from a numerical overall normalization factor, 
the same form as the original one (\ref{partitionfunction}) 
with the interchange 
$P\leftrightarrow Q, \, B \leftrightarrow \tilde{B}$ 
\begin{equation}
 Z=\bigl(
\prod_{y, \mu}\,  \sum_{\tilde{n}_y, \tilde{m}_{y,\mu}}\bigr)
\prod_y \delta^{(P)}
(\triangle_{\mu} \tilde{m}_{y, \nu}-\triangle_{\nu}
\tilde{m}_{y, \mu})
\prod_{y, \mu}
\tilde{B}(\triangle_{\mu}\tilde{n}_y - Q \tilde{m}_{y, \mu}).
\end{equation}
The global modes represented by $\tilde{m}_{y, \mu}$ are 
interpreted as the analogue of the momentum mode of 
toroidally compactified strings, and hence the 
dual transformation interchanges the winding and 
momentum modes, precisely as required for  
T-duality symmetry.   
In particular, the systems with $P=Q$ 
are self-dual on surfaces of arbitrary genus.  

Note that, although the 
models with a given $K=PQ$ are equivalent 
to the usual $Z_K$ spin models on the 
sphere, they are  in general different on higher-genus surfaces, 
because of the different appearance of the  
global momentum and winding modes, 
depending on the choice of $P$ and $Q$. 
In a sense, the $Z_{P,Q}$ model amounts to {\it compactifying} the 
$Z_K$ target space by a subgroup $Z_P$ 
of the {\it isometry} group $Z_K$. 
For example, the standard Ising model corresponds to 
$P=2, Q=1$, $Z_{2,1}$ model.  
Hence its dual is $Z_{1,2}$. Therefore the Ising model is 
not exactly self-dual on higher genus, as discussed in \cite{akosy}.  
In this class of models, the simplest exactly self-dual model is 
$Z_{2,2}$, which is identical with 
the ($Z_4$) Ashkin-Teller model on the sphere. 
Finally, we note that taking the limit 
$P=Q \rightarrow \infty$ appropriately 
gives a lattice version of  toroidally compactified 
strings discussed in the 
Appendix A of \cite{akosy}. 
A brief discussion on this limit will be given in 
section IV.

\section{Self-dual matrix models: General Theory}
\setcounter{equation}{0}

We now show how to construct the matrix model corresponding to the 
$Z_{P,Q}$ model.  In the standard method of coupling the $Z_K$ spin 
system to 2D gravity, we introduce $K$
 different Hermitian $N\times N$ 
matrices $M_a, \, \, a =1, 2, \ldots, K$ and assume the action 
\begin{equation}
S = N{\rm Tr}\left[{1\over 2}\sum_{a, b=1}^KC_{ab}M_a M_b+ 
\sum_{a=1}^K V(M_a)\right]. 
\end{equation}
In the Feynman graph expansion, the potential $V(M_a)$ 
represents a spin site which is on the 
center of a discretized surface element.  The propagator 
$C^{-1}_{ab}$, on the other hand, corresponds to 
the Boltzmann factor assigned to a link connecting 
the nearest neighbor spin sites 
with spin variables $a$ and $b$.  
The $Z_K$ symmetry requires that 
the kinetic operator (and the propagator)  
satisfies translation invariance modulo $K$ with respect to the 
indices $a, b$,  
$$
C_{ab}= C(a-b)=C(a-b \pm K), \,  \, C^{-1}_{ab}\equiv D(a-b)=D(a-b  \pm K). 
$$

In our terminology, this construction only represents 
the $Z_{K,1}$ model. To extend the construction to 
general $Z_{P,Q}$ matrix models, 
we first  introduce a set of 
$P$ different Hermitian $QN\times QN$ matrices, 
denoted as $M_a, \, \, a=1,2, \ldots, P$. 
The index $a$ ranging from $1$ to $P$ corresponds 
to the $Z_P$ spin variables as above, 
while the additional $Q\times Q$ matrix elements of each $M_a$ 
are supposed to be associated with the $Z_Q$ link variables. 
To prevent unnecessary complication in notations,  
we will always suppress the original $U(N)$ indices 
in the following  
and only indicate the additional $Q\times Q$ matrix indices by 
$i, j, \ldots \in \{1, 2, \ldots, Q\}$. 
The choice of an appropriate propagator will then enable us to 
construct the $Z_{P,Q}$ matrix model.  The correct choice is, 
suppressing the $U(N)$ indices, as follows:  
\begin{equation}
D^{a, b}_{ ij, kl}  ={1\over N} \sum_{m=0}^{Q-1}B(a-b-Pm)(L^m)_{il}
(L^{\dagger \, m})_{kj}, 
\label{propagator}
\end{equation}
where $B$ is nothing but the Boltzmann factor 
used in the previous section, and 
$L$ is an arbitrary $Q\times Q$ unitary matrix 
satisfying the conditions
\begin{equation}
L^Q = I , \, \, L^{-1} = L^{\dagger},
\label{pcondition1}
\end{equation}
\begin{equation}
{\rm tr}_Q L^i= 0  \, \,  \, {\rm if} \, \, i\ne 0 \, \,  \, 
({\rm modulo}  \, \, Q).
\label{pcondition2}
\end{equation}
Here the trace notation ${\rm tr}_Q$ 
means taking the trace only with respect to 
 the indices $i,j, \ldots$. 
The notation ${\rm Tr}$ will be used for the total trace operation 
including both the $U(N)$ and the 
$U(Q)$ indices.  On the other hand, the 
trace operation with respect only to the ordinary $U(N)$ indices 
will be denoted 
by ${\rm tr}_N$. 
A formula for the inverse of the propagator 
(\ref{propagator}) 
will be given later. 

The potential function corresponding to 
the spin sites are assumed to be the following 
special form using the result of \cite{siegel}
\begin{equation}
N{\rm Tr} \sum_{a=1}^P \ln (1- gM_a). 
\end{equation}
This is necessary to ensure that the measure for the random 
triangulation is invariant under the T-duality transformation 
even before taking the (double) scaling limit. 

The model is invariant under the group $U(N)\times U(Q), 
M_a \rightarrow UM_aU^{-1}, \, \, U\in U(N)\times U(Q)$,  
provided the matrix $L$ is transformed by the $U(Q)$ part of $U$. 
It is easy to check that the conditions (\ref{pcondition1}), 
(\ref{pcondition2}) 
precisely impose the constraint corresponding to the 
$\delta^{(Q)}$ function in (\ref{partitionfunction}) for general 
triangulation of arbitrary surfaces, at each 
elementary closed circle of links (i.e., plaquette). 
Hence, the matrix $L$ is arbitrary under the 
$U(N)\times U(Q)$ invariant conditions (\ref{pcondition1}), 
(\ref{pcondition2}).   
Also it is not difficult to prove explicitly 
that this model restricted to the sphere approximation 
is equivalent to the standard $Z_K$ matrix model 
as briefly discussed in section IV. 

Now that we have given the definitions of the matrix models, let 
us next proceed to rewriting of the models 
in a manifestly T-duality 
symmetric form.  We will extend the method given in 
\cite{siegel} for the case  of  pure gravity to our 
$Z_{P,Q}$ model. For this purpose, it is 
convenient to go to a particular representation 
for the matrix $L$ using the $U(Q)$ symmetry, 
namely the diagonalized representation given by 
\begin{equation}
L_{ij}=\delta_{ij}{\rm e}^{i{2\pi \over Q}(i-1)}.
\end{equation} 
Thus the propagator is now 
\begin{equation}
 D^{ac, bd}_{ ij, kl}  ={1\over N}\delta_{ac}
\delta_{bd} \sum_{m=0}^{Q-1}B(a-b-Pm)
\delta_{il}\delta_{kj}{\rm e}^{
i{2\pi\over Q}(i-j)m}.
\label{propagator2}
\end{equation}
Here, by putting additional Kronecker $\delta$'s, 
$\delta_{ac}, \delta_{bd}$, 
we duplicated the indices $a$ and $b$ 
in order to make the appearance of the $i, k \, (\in Z_Q)$ 
and $a, b \,(\in Z_P)$ indices symmetric.  
We note that in this representation the propagator has 
manifest $Z_Q$ 
periodicity with respect to the indices $i,j,\ldots$ 
while the periodicity with respect to the indices $a, b, \dots$ 
is not manifest. 
This is expected from the situation of the fixed-surface model 
of the previous section, since the $Z_P$ periodicity appears only 
after summing over the link variables.  In the matrix model, 
the latter operation appears only for Feynman amplitudes 
after taking the trace operation.  

We first  introduce $PN$ auxiliary 
vector fields $\psi_i^a $ which 
transform as a complex vector
under the group $U(N)\times U(Q)$.  
Here as before we have suppressed the $U(N)$ 
vector (``color") indices and 
also the ``flavor" $U(N)$ vector indices, while 
both of the matter indices $a (\in U(P)$ vector)
 and $i (\in U(Q)$ vector)  are explicitly 
indicated. Thus the 
auxiliary fields $\psi_i^a$ can  actually be treated as 
$NQ\times NP$ matrix fields which transform as 
$$
\psi_i^a \rightarrow (U_c\psi V_f)_i^a ,
$$
where $U_c\in U(N)\times U(Q)$ and $V_f \in U(N)\times U(P)$.  
Using these auxiliary fields, rewrite the potential term as 
\begin{equation}
{\rm e}^{-N{\rm tr}_N\sum_{a=1}^P\ln(1-gM_a)}
=\int d\psi d\psi^{\dagger} \, 
{\rm e}^{-\sum_{a=1}^P\sum_{i,j=1}^Q {\rm tr}_N
\psi^{\dagger\,a}_i(1-gM_a)_{ij}\psi_j^a}.
\end{equation}
Here we used the notation $\psi^{\dagger}$ for the 
complex conjugate of the $\psi$'s 
treating them as $N\times N$ (color $\times$ flavor) matrices. 
We can then perform the integration over the matrix $M_a$. 
Using the propagator (\ref{propagator2}), 
the partition function  
takes the form 
\begin{equation}
Z={\cal N}_{P,Q}(B) \int d\psi d\psi^{\dagger}\, \exp\left[
-\sum_{i=1}^Q\sum_{a=1}^P{\rm tr}_N \psi_i^a \psi^{\dagger\, a}_i
+{1\over 2}
g^2 \sum_{i,j,k,l=1}^Q\sum_{a,b,c,d=1}^P {\rm tr}_N
 \psi_j^b \psi^{\dagger\, a}_iD_{ij, kl}^{ab,cd}
\psi_l^d\psi^{\dagger\, c}_k
\right].
\end{equation}
The normalization constant ${\cal N}_{P,Q}(B)$ originates from the 
Gaussian integration,
$$
{\cal N}_{P,Q}(B)={\rm det}^{1/2}_{(NPQ)^2}D ,
$$ 
where $D$ is the matrix of the propagator (\ref{propagator2}). 
 
The T-dual transformation amounts to rewriting the partition function 
in terms of the dual Boltzmann factor $\tilde{B}(a)$, defined by 
(\ref{fouriertransboltzmann}). 
By directly substituting (\ref{fouriertransboltzmann}) to the 
propagator (\ref{propagator2}), we find 
\begin{equation}
D_{ij,kl}^{ab,cd}=
{Q\over N\sum_{b=0}^{K-1}\tilde{B}(b)}\sum_{\tilde{m}=0}^{P-1}
\tilde{B}(i-j-Q\tilde{m}) {\rm e}^{
i{2\pi \over K}(a-c)(i-j-Q\tilde{m})}
 \delta_{ab}\delta_{cd}
\delta_{il}\delta_{jk}.
\label{dform}
\end{equation}
A little examination of this expression
 shows that the partition function in terms of the $\tilde B$ 
 takes the same general form 
as the original one apart from the normalization factor, 
\begin{equation}
Z={\cal N}_{P,Q}(B) \int d\phi d\phi^{\dagger}\, \exp\left[
-\sum_{a=1}^P\sum_{i=1}^Q{\rm tr}_N  \phi_a^i \phi^{\dagger\, i}_a
+{1\over 2}
\tilde{g}^2 \sum_{a,b,c,d=1}^P \sum_{i,j,k,l=1}^Q{\rm tr}_N
\phi_a^i \phi^{\dagger\, l}_d
\tilde{D}_{da, bc}^{li,jk}\phi_c^k \phi^{\dagger\, j}_b
\right], 
\end{equation}
after interchanging the $Z_P$ and $Z_Q$ indices by 
making the following redefinitions of the auxiliary vectors, propagator 
and the coupling constant, respectively, as 
\begin{equation}
\phi_a^i \equiv {\rm e}^{i{2\pi\over K}ai}\psi^{\dagger\, a}_i ,
\label{phaseab}
\label{redefaux}
\end{equation}
\begin{equation}
\tilde{D}_{da, bc}^{li,jk}
\equiv {1\over N}\sum_{\tilde{m}=0}^{P-1}\tilde{B}(l-j-Q\tilde{m})
{\rm e}^{i{2\pi\over P}(d-a)\tilde{m}}
\delta_{ab}\delta_{cd}\delta_{li}\delta_{jk} ,
\end{equation}
\begin{equation}
g^2\rightarrow \tilde{g}^2={g^2Q\over \sum_{b=0}^{K-1}\tilde{B}(b)}
={g^2\sum_{a=0}^{K-1}B(a)\over P}.
\label{redefineg}
\end{equation} 
Here  we have   
interchanged the color and flavor indices by performing a 
cyclic permutation for the auxiliary fields $\psi$'s 
in the quartic term of the action
\footnote{
Note that we have used the same trace notation ${\rm tr}_N$ 
for $U(N)$ flavor indices. 
}, 
corresponding to the replacement, 
$$
D_{ij, kl}^{ab, cd}(B) \rightarrow {{\rm e}}^{i{2\pi \over K}(a-c)(i-j)}
\tilde{D}_{da, bc}^{li, jk}(\tilde{B}), 
$$
of which the phase factor is absorbed by (\ref{phaseab}). 
We note that in this dual form, the $Z_P$ periodicity 
instead of the $Z_Q$ periodicity is now manifest. This comes about 
due to the redefinition  (\ref{redefaux}) of the auxiliary fields: 
The field $\psi_i^a$ is supposed to be manifestly $Z_Q$ 
periodic with respect to the index $i$, while it  acquires a 
phase ${\rm e}^{i{2\pi\over Q}i}$ under the translation 
$a\rightarrow a+P$. This property is interchanged 
after the above redefinition from $\psi$ to $\phi$. 
It should also be remarked that 
 the auxiliary fields $\psi$ and $\phi$ carry 
both the matter  ($Z_P$) and dual-matter ($Z_Q$) 
indices in addition to the 
color and flavor $U(N)$ indices. This is an interesting aspect of the 
present formalism, which is something not appeared in the 
familiar formulation of  T-duality in string theory. 

The above result 
is precisely the dual 
transformed structure of the $Z_{P,Q}$ models, 
apart from the redefinition of the coupling constant which is 
connected to the overall normalization factor suppressed 
 in the previous discussion for the fixed surface. 
It is evident that 
by reversing the route from the $M_a \, \, \, (a=1, 2, \ldots, P)$ 
representation to the $\psi_i^a$ representation, we can 
construct the dual transformed matrix model, now with 
$Q$ Hermitian $NP\times NP$ 
matrices $\tilde M_i \, \, \, (i=1, 2, 
\ldots, Q)$ from the $\phi_a^i$ representation. 
Thus we have established that our matrix models have desired 
T-duality properties.  
Formally, 
the T-duality symmetry corresponds to the 
following identity for the partition function,
\begin{equation}
{Z_{P,Q}(B, g)\over {\cal N}_{P,Q}(B)}
={Z_{Q,P}(\tilde{B}, \tilde{g})\over {\cal N}_{Q,P}(\tilde{B})}.
\end{equation}
Note that the normalization constants 
${\cal N}_{P,Q}$ and ${\cal N}_{Q,P}$  whose ratio 
is easily calculable using (\ref{dform}) 
 are independent of 
the coupling constant $g$. As a consequence, the 
normalization constants can be neglected in the 
scaling limit.

In order to write down the actions of the matrix models 
explicitly in terms of the Boltzmann factor 
$B$, we need a general formula 
for the inverse of the propagator 
(\ref{propagator}). This is easily obtained by applying $Z_P$ 
Fourier transformation appropriately. The result is 
\begin{eqnarray}
C_{ij, kl}^{ab}&\equiv&(D^{-1})_{ij,kl}^{ab} \nonumber \\
&=& N\delta_{il}\delta_{jk}{1\over P}{\rm e}^{-
i{2\pi\over K}(a-b)(i-k)}
\sum_{m=0}^{P-1}{1\over \tilde{B}(m;i-k)}
{\rm e}^{i{2\pi\over P}(a-b)m} ,
\end{eqnarray}
where
\begin{equation}
\tilde{B}(m;i-k)=\sum_{a=0}^{P-1}\sum_{l=0}^{Q-1} B(a-Pl) 
{\rm e}^{-i{2\pi\over K}(a-Pl)(i-k)} {\rm e}^{-i{2\pi\over P}ma}.
\end{equation}
Corresponding to the $Z_Q$ periodicity of the propagator, this 
form of the kinetic term is $Z_Q$ periodic with respect to 
the indices $i, j, \ldots$,
 while under the $Z_P$ translation it 
is periodic only up to a phase 
${\rm e}^{-i{2\pi\over Q}(i-k)}$. 
Of course, by inverting this formula, we could have started from the 
general form of the kinetic term having the 
same periodicity and expressed the Boltzmann factor 
in terms of the kinetic term. 

Finally, we briefly touch upon the question of 
observables in our models. 
Since we have to preserve the $U(N)\times U(Q)$ symmetry, 
the set of the most general invariants consists 
of the traces of 
arbitrary polynomials consisting 
of the matrices $M_a$ and ${\cal L}$ 
where ${\cal L}$ is the $U(N)\times U(Q)$ matrix acting as the identity 
in $U(N)$ and as $L$ in $U(Q)$: 
$$
A_{ab\cdots}\,(n,l,\ldots;m,p,\ldots) = 
{1\over NQ}{\rm Tr} (M_a^n{\cal L}^m M_b^l {\cal L}^p \cdots).
$$
If the total power of the matrix ${\cal L}$ is $q \, \,$ (mod $Q$), 
$A_{ab\cdots}\,(n,l,\ldots;m,p,\ldots)$ represents a loop state with 
``winding number $q$".  
In the standard models, it has been impossible to explicitly 
extract the winding modes in terms of the matrix variables. 
Our construction thus indicates how to remedy this 
deficiency of the usual matrix models by extending the $U(N)$ symmetry 
to $U(N)\times U(Q)$.

\section{Examples and Discussions}
\setcounter{equation}{0}

The purpose of this section is to first present a 
few simple special cases of our extended models in  more 
concrete forms as an illustration 
of the general theory, and then further discuss some of the  
important aspects of the models, in 
particular,  the reduction of the degrees of freedom and 
the $c=1$ limit.   
 
\subsection{Examples}
We give concrete formulas  for three cases, 
 $Z_{1,2}$, $Z_{1,3}$ and $Z_{2,2}$. 
 
\vspace{0.3cm}
\noindent
(i) Duality between $Z_{1,2}$ and $Z_{2,1}$ matrix models

This corresponds to  the case treated in \cite{caletal,akosy} 
with cubic 
potential. 
The $Z_{1,2}$ model has only link variables.  
The dual version, on the other hand, is the $Z_{2,1}$ model 
which is nothing but the ordinary two-matrix model with 
spin variables, but without any link variables. 
Let us confirm this by an explicit computation.  
The propagator of the $Z_{1.2}$ matrix model is written as 
\begin{equation}
\langle M_{ij}M_{kl}\rangle=\frac{1}{N}(\delta_{il}\delta_{kj}
+{\rm e}^{-2\beta}L_{il}L_{kj}),
\end{equation}
where $\beta$ is an inverse temperature of the $Z_{1,2}$ spin system.
$L$ has the properties in  Eqs. 
({\ref{pcondition1}) and ({\ref{pcondition2}) with $Q=2$. Correspondingly, 
the action is 
 \begin{equation}
S_M =  N\frac{1}{1-{\rm e}^{-4\beta}}\frac{1}{2}
{\rm Tr}(M^2-{\rm e}^{-2\beta}
{\cal L}M{\cal L}M)+N{\rm Tr}\ln (1-gM).
\end{equation}
Introducing the auxiliary 
variables $\psi_i$ and performing the $M$-integral,
 we have the action 
\begin{equation}
S_{\psi} = N\sum_{i=1}^2{\rm tr}_N\psi^{\dagger}_i\psi_i
   -N\frac{1}{2}g^2{\rm tr}_N\left\{
(\sum_{i=1}^2\psi^{\dagger}_i\psi_i)^2+{\rm e}^{-2\beta}(\sum_{i,j=1}^2
\psi^{\dagger}_iL_{ij}\psi_j)^2\right\}.
\end{equation}
For the purpose of  direct transition to the 
dual-matrix representation in the simple example 
treated here, it is convenient to 
 introduce two  auxiliary matrices $U$ and $V$, 
which are Hermitian and have 
$U(N)$ flavor indices, 
instead of faithfully following the procedure 
adopted in the general theory. 
Then the action $S_{\psi}$ can be rewritten as 
\begin{equation}
S_{UV\psi} =  N{\rm tr}_N\psi^{\dagger}_i\psi_i
+N\frac{1}{2}{\rm tr}_N(U^2+{\rm e}^{2\beta}V^2) 
-Ng{\rm tr}_N(\sum_i \psi_iU\psi_i^{\dagger}
+\sum_{i,j}L_{ij}\psi_jV\psi^{\dagger}_i).
\end{equation}
Integrating $\psi$ and changing the basis as 
\begin{equation}
\tilde{M}_1=U+V,\,\,\tilde{M}_2=U-V,
\label{changeUV}
\end{equation}
we get the dual action 
\begin{equation}
\tilde{S}_M=N\frac{1}{1-{\rm e}^{-4\tilde{\beta}}}\frac{1}{2}{\rm tr}_N
\left[\tilde{M}_1^2
+\tilde{M}_2^2-2{\rm e}^{-2\tilde{\beta}}\tilde{M}_1\tilde{M}_2\right]
+N\sum_{a=1}^2{\rm tr}_N\ln (1-\tilde{g}\tilde{M}_a),
\label{Z12dual}
\end{equation}
where we rescaled $\tilde{M}_a$ as 
$\tilde{M}_a\rightarrow\sqrt{1+{\rm e}^{-2\beta}}\tilde{M}_a$ 
corresponding to the redefinition (\ref{redefineg}). 
The dual temperature $\tilde{\beta}$ is defined by the Fourier 
transformation (\ref{fouriertransboltzmann}) as 
$$
{\rm e}^{-2\tilde{\beta}}=\tanh \beta,
$$
which is of course the famous Kramers-Wannier relation.

\vspace{0.3cm}

\noindent
(ii) $Z_{1,3}$ matrix model

We repeat a similar calculation for the $Z_{1,3}$ matrix model. 
The propagator and the action have the following forms:
\begin{equation}
\langle M_{ij}M_{kl}\rangle = \frac{1}{N}[\delta_{il}
\delta_{kj}+{\rm e}^{-\frac{3}{2}\beta}L_{il}(L^2)_{kj}
+{\rm e}^{-\frac{3}{2}\beta}(L^2)_{il}L_{kj}], 
\end{equation}
\begin{equation}
S_M =  N\frac{\coth\frac{3}{4}\beta}{1+2{\rm e}^{-\frac{3}{2}\beta}}
\frac{1}{2}{\rm Tr}\left[M^2
-\frac{2}{1+{\rm e}^{\frac{3}{2}\beta}}{\cal L}M{\cal L}^2M\right]
+N{\rm Tr}\ln (1-gM).
\end{equation}
Through the same steps as before, we obtain the action $S_{\psi}$ of the 
variables $\psi_i$'s. Then, 
introducing a Hermitian matrix $U$ and 
a complex matrix $X$, we rewrite $S_{\psi}$ as 
\begin{eqnarray}
S_{UX\psi} & = & N\sum_{i=1}^3{\rm tr}_N\psi^{\dagger}_i\psi_i
+N\frac{1}{2}{\rm tr}_N(U^2+
2{\rm e}^{\frac{3}{2}\beta}X^{\dagger}X) \nonumber \\
 & & -Ng\sum_{i=1}^3{\rm tr}_N(\psi_iU\psi^{\dagger}_i)
-Ng\sum_{i,j=1}^3{\rm tr}_N(X^{\dagger}\psi^{\dagger}_iL_{ij}\psi_j+
\psi^{\dagger}_i(L^2)_{ij}\psi_jX).
\end{eqnarray}
After the $\psi$-integral and the replacement 
\begin{equation}
\tilde{M}_1=U+X+X^{\dagger},\,\,\tilde{M}_2
=U+\omega^2X+\omega X^{\dagger},\,\,
\tilde{M}_3=U+\omega X+\omega^2X^{\dagger},
\end{equation}
with $\omega={\rm e}^{i\frac{2\pi}{3}},$ we obtain the action of the 
$Z_{3,1}$ matrix model 
 \begin{eqnarray}
\tilde{S}_M & = & 
N\frac{\coth\frac{3}{4}\tilde{\beta}}{1+2{\rm e}^{-\frac{3}{2}
\tilde{\beta}}}\frac{1}{2}{\rm tr}_N\left[\sum_{a=1}^3\tilde{M}_a^2
-\frac{2}{1+{\rm e}^{\frac{3}{2}\tilde{\beta}}}(\tilde{M}_1\tilde{M}_2
+\tilde{M}_2\tilde{M}_3+\tilde{M}_3\tilde{M}_1)\right] \nonumber \\
 & & +N\sum_{a=1}^3{\rm tr}_N\ln (1-\tilde{g}\tilde{M}_a),
\end{eqnarray}
where the rescaling 
$\tilde{M}_a\rightarrow\sqrt{1+2{\rm e}^{-\frac{3}{2}\beta}}\tilde{M}_a$ 
was done, and the dual temperature is given by 
$$
{\rm e}^{\frac{3}{2}\tilde{\beta}}=
\frac{{\rm e}^{\frac{3}{2}\beta}+2}{{\rm e}^{\frac{3}{2}\beta}-1}.
$$
\vspace{0.3cm}

\noindent
(iii) $Z_{2,2}$ matrix model 

This is the simplest example of self-dual models. 
If we adopt the standard Boltzmann factor for the 
$Z_4$ model,
$$
B(\Delta_{\mu}n_x-2m_{x,\mu})={\rm e}^{\beta[\cos\frac{2\pi}{4}
(\Delta_{\mu}n_x-2m_{x,\mu})-1]},
$$
having a self-dual structure with 
\begin{equation}
{\rm e}^{\tilde{\beta}}=\coth\frac{\beta}{2}, 
\label{dualtemp3}
\end{equation}
 the 
propagator 
and the corresponding action are  given, respectively, as
\begin{eqnarray}
\langle (M_1)_{ij}(M_1)_{kl}\rangle & = & \langle (M_2)_{ij}(M_2)_{kl}\rangle 
=\frac{1}{N}(\delta_{il}\delta_{kj}
+{\rm e}^{-2\beta}L_{il}L_{kj}), \nonumber\\
\langle (M_1)_{ij}(M_2)_{kl}\rangle & = & \frac{1}{N}{\rm e}^{-\beta}
(\delta_{il}\delta_{kj}+L_{il}L_{kj}), 
\end{eqnarray}
\begin{eqnarray}
S_M & = & N\frac{{\rm e}^{2\beta}}{4\sinh^2\beta}\frac{1}{2}{\rm Tr}
\left\{\sum_{a=1}^2[(M_a)^2
+{\rm e}^{-2\beta}{\cal L}M_a{\cal L}M_a]-2{\rm e}^{-\beta}M_1M_2
-2{\rm e}^{-\beta}{\cal L}M_1{\cal L}M_2\right\}
\nonumber \\
 & & +N\sum_{a=1}^2{\rm Tr}\ln [1-gM_a].
\end{eqnarray}
Following the steps of the general theory, we can easily confirm that the 
model is self-dual with respect to (\ref{dualtemp3}) 
and $\tilde{g}=g\frac{1+{\rm e}^{-\beta}}{\sqrt{2}}.$

 \subsection{Reduction of degrees of freedom}

      From the discussion of section II, it is clear that the 
$Z_{P,Q}$ matrix models must be equivalent 
to the standard $Z_K$ matrix
models when restricted to the sphere approximation.
Apparently, however, 
the $Z_{P,Q}$ matrix models have more degrees of freedom
than the $Z_K$ models with $K = PQ$. If counted as $N\times N$ 
Hermitian matrices, we have $PQ^2$ matrices instead of $PQ$.  
Let us therefore examine how the reduction of degrees of freedom 
occurs.  It will turn out that  
there is a sort of local gauge symmetry in Feynman
diagrams which accounts for the superfluous degrees of freedom.

For simplicity, we here consider a concrete example of 
the $Z_{P,2}$ ($Q=2$) models.  
Once we understand this special case, 
extension to the general case will be straightforward. 
For this purpose, it is more 
advantageous to use the real representation of 
the matrix $L$ given by
\begin{equation}
L_{ij} = \delta^{(Q)}_{i+1 \, j},
\end{equation}
than the diagonal representation used in the 
general theory in section III. 
 The nonzero components of the propagator then satisfy the 
following symmetry properties with respect to the $Z_2$ indices ($i,j$ type) 
\begin{eqnarray}
D_{11,11}&=&D_{22,22}=D_{12, 21}=D_{21,12}, \nonumber \\
D_{11,22}&=&D_{22,11}=D_{12,12}=D_{21,21},
\label{d-identity}
\end{eqnarray}
where we have suppressed the $Z_P$ indices  ($a,b$ type) since 
the identities are valid for arbitrary $a, b$. 
All other components of the propagator vanish. 
In particular, there is no coupling between the diagonal 
and off-diagonal matrix elements with respect to the $Z_2$ indices. 
This shows that the links assigned 
to the off-diagonal matrix elements 
must always form $Z_2$ closed loops. 
Furthermore, an amplitude with the off-diagonal elements 
is identical with a corresponding diagram in which all the 
off-diagonal matrix elements are replaced by the corresponding 
diagonal elements as indicated in (\ref{d-identity}). 
Note here that the vertices behave as identities 
with respect the $Z_2$ indices and hence do not have 
any contribution violating the symmetry between the 
diagrams with diagonal and off-diagonal elements. 
In the case of sphere topology, the correspondence between 
the diagrams is established by making the interchange
$1 \leftrightarrow 2$ globally 
for all $Z_2$ indices inside the domain enclosed 
by the closed curve of off-diagonal links. 
This replaces the off-diagonal links along the closed curve to links with 
diagonal matrix elements, without changing the amplitude. 
 We can start 
this procedure from the smallest 
domains and go to larger ones successively  
to eliminate all off-diagonal links in this way.  
Thus the Feynman diagram contributions have a kind of 
gauge degeneracy  $2^L$ 
which is determined by the number $L$ of independent 
$Z_2$ closed loops of off-diagonal links on the surface. 
Using the fact that $L$ coincides 
with the number of dual sites, 
it is easy to see 
that the free energy 
becomes essentially identical 
with the model with only the diagonal matrix 
elements $(M_a)_{ii}, \, (i=1,2)$ under the rescaling of $N$ and the 
coupling constant as $N \rightarrow 2N, g_n\rightarrow  2^{{n-2\over 2}}g_n$ 
where $g_n$ is the coupling constant for the term 
${g_n\over n}{\rm Tr}M_a^n$ in the potential.  
Thus the matrix spin degree of freedom is reduced to $2P$ which 
is the same as for the standard $Z_{2P}$. Obviously, the 
present argument cannot be extended to higher genus, since 
the above correspondence between diagonal and 
off-diagonal amplitudes cannot be 
established when the closed loop 
of off-diagonal links wraps around a non-trivial homology cycle. 

Extension to the general case is straightforward. 
The links with off-diagonal elements form $Z_Q$ closed loops and 
the propagator satisfies the identities
$ D_{ij,kl} = D_{i+m \, j+n, k+n \, l+m}$. 
Repeating the above arguments, we arrive at the 
rescaling $N \to QN, \quad g_n \to Q^{\frac{n-2}{2}}g_n$ 
after the system is reduced to matrices with diagonal (with 
respect to the $Z_Q$ indices) elements.  
The same result can also be obtained by a simple counting of  the 
Feynman diagrams by focusing on their dependence with respect to
$N, Q$ and the coupling constant.

\subsection{$c=1$ limit}

We next turn to the problem of the 
$c=1$ limit of our models which corresponds to 
taking the limit $P=Q\rightarrow \infty$. This problem 
is important if one regards the present 
matrix models as a nonperturbative regularization of the 
T-duality symmetry of string theory. 
 
First of all, take the  $Z_{Q,Q}$ model assuming  
the most general form of 
the Boltzmann factor \cite{alcarazcardy} as 
$$
B(\alpha)=\exp \left[\sum_{\delta=0}^{Q^2-1} K_{\delta} 
          \left(\cos \frac{2\pi \alpha \delta}{Q^2} -1 \right)
\right],
$$
where $K_{\delta}$'s are constants.  
In the limit $Q\rightarrow \infty$, we obtain the standard 
Gaussian Boltzmann factor as follows:  
when $K_{\delta}$ is large  
with $\frac{K_{\delta}}{Q^2}=\beta_{\delta}$ being fixed, 
we can expand the cosine to obtain
\begin{eqnarray}
B(\alpha) & \sim & 
\prod_{\delta=0}^{Q^2-1} \sum_{n_{\mu}^{(\delta)} \in Z}
\exp\left[-\frac{\beta_{\delta}}{2} \delta^2 
     (\frac{2 \pi}{Q}\alpha - 2\pi Qn_{\mu}^{(\delta)})^2
\right] 
\nonumber \\
          & = & \sum_{n_{\mu}\in Z} \exp\left[-\frac{R^2}{4 \pi}
               (\frac{2\pi}{Q}\alpha-2\pi Qn_{\mu})^2\right].
\label{gaussian}
\end{eqnarray}
In the last expression, we left one of $\beta_{\delta}$'s 
with $\delta=O(1)$ nonzero 
and redefined
$\frac{R^2}{2\pi} \equiv \beta_{\delta} \delta^2$ 
to make $B(\alpha)$  
fit to the ordinary Gaussian Boltzmann factor of the $c=1$ model. 

For taking the $c=1$ limit on a  fixed lattice, 
we make the following replacement
$$ 
\frac{2\pi n}{Q} \rightarrow X,~ 
\frac{2\pi}{Q}\sum_{n \in Z_Q} \rightarrow \int_0^{2\pi} dX.
$$
Then the partition function of the $Z_{Q,Q}$ model  becomes, 
apart from a numerical proportional factor,            
\begin{equation}
Z = \int_0^{2\pi} dX \sum_{N_{\mu} \in Z} 
    \delta_{\epsilon_{\mu \nu} \Delta_{\mu} N_{\nu}}
    \exp\left[- \frac{R^2}{4\pi}(\Delta_{\mu}X-2\pi N_{\mu})^2
\right],
\label{zpp}
\end{equation}
where we rewrite $m_{\mu}+Qn_{\mu}$ by $N_{\mu}$. 
Here and below, we suppress the subscripts for sites and 
correspondingly the product symbols for sites for notational 
simplicity.   
In the dual representation, corresponding to 
$\tilde{n} \in Z_Q$ and $\psi \in Z_K$ 
in section II, it is natural to take the limit as,
$$ 
\frac{2\pi \tilde{n}}{Q} \rightarrow \tilde{X},~ 
\frac{2\pi}{Q}\sum_{\tilde{n} \in Z_Q} 
\rightarrow \int_0^{2\pi} d\tilde{X},~
\frac{2\pi}{Q}\sum_{\psi_{\mu} \in Z_K} 
\rightarrow \int_{-\infty}^{\infty} d\psi_{\mu}.
$$
Similarly to the section II, the summation 
over $N_{\mu}$ leads to 
$
\psi_{\mu}=\Delta_{\mu}\tilde{X}-2\pi \tilde{m}_{\mu},~
\tilde{m}_{\mu} \in Z,
$ 
and the summation over $m_{\mu}$ imposes the constraint 
on $\tilde{m}_{\mu}$. 
Thus the partition function after the dual transformation 
takes the same form as (\ref{zpp}),
\begin{equation}                                              
Z \propto 
  \int_0^{2\pi} d\tilde{X} \sum_{\tilde{m}_{\mu} \in Z} 
  \delta_{\epsilon_{\mu \nu} \Delta_{\mu} \tilde{m}_{\nu}}
  \exp\left[- \frac{1}{4\pi R^2}
       (\Delta_{\mu}\tilde{X}-2\pi \tilde{m}_{\mu})^2
\right],
\label{dualzpp}
\end{equation}
with the correspondence $R \leftrightarrow 1/R$.
As has been discussed in \cite{gross-kleb} and 
 the Appendix A of  \cite {akosy}  
it is also straightforward to show self-duality 
by a direct dual transformation 
 (\ref{zpp}) $\leftrightarrow$ (\ref{dualzpp})  
in the continuous target space.   

Let us next consider the $Q \rightarrow \infty$ limit 
of the matrix model corresponding to the $Z_{Q,Q}$ model. 
Since the model is manifestly 
self-dual for an arbitrary integer $Q$, 
 all we need to check is 
whether the dual transformation laws of physical parameters 
are well-defined in the limit.  
There are two such parameters, 
namely, the compactification radius $R$ 
and the coupling constant $g$. 
First, as for $R$, it follows from (\ref{zpp}) 
and (\ref{dualzpp})   
that $R$ is transformed              
as $R \leftrightarrow 1/R$, since the transformation law of the 
Boltzmann factor is of course identical with the case of 
the fixed lattice.  The transformation of the 
coupling constant is given by (\ref{redefineg})  which 
in the present limit leads to 
\begin{equation}
g^2 \rightarrow \tilde{g}^2 = {g^2 \over R}. 
\label{dualrelation}
\end{equation}
Thus both are well-defined in the $c=1$ limit.
 
In particular, nonsingular transformation 
property of the coupling constant 
would allow a well-defined double scaling limit 
in the sense of world sheet. From the relation (\ref{dualrelation}), 
we expect that the 
critical point $g_c$  as a function of 
$R$ would behave as 
$$
g_c= f(R)R^{1/4},
$$
where $f(R)$ is a function invariant under the dual transformation 
$R\leftrightarrow 1/R$. 
Unfortunately,  however, self-duality alone is 
not powerful enough for obtaining 
exact results in the scaling limit.  For example, it is possible  
that the function $f(R)$ actually vanishes in the 
$c=1$ limit due to the degeneracy, 
as suggested from the counting argument 
of the previous subsection. In that case, 
the interplay between the $c=1$ 
limit  and the scaling limit is rather subtle, 
since  taking the formal zero-coupling limit $g\rightarrow 0$  
directly in the action 
is not meaningful and we have to examine the 
physical correlators   
as in the usual situation in general lattice field theories. 
In any case, however, our discussion is sufficient to show that 
the present models can serve as a nonperturbative regularization 
of the T-duality symmetry 
in the continuum critical string theories.  
In particular, it is guaranteed that the  properly defined 
continuum partition 
function is duality symmetric, as 
has been explicitly computed \cite{gross-kleb} 
in the standard $c=1$ model 
of finite radius.

\subsection{Conclusion}
We have established a new class of extended matrix models 
which have manifest symmetry under the T-duality 
transformation. We can think of several 
possible extensions and applications of our work. 
Among others, we would like  to mention the 
following.
\begin{enumerate}
\item Construction of manifestly T-duality symmetric 
string field theories: Using the macroscopic loop 
variables including the ${\cal L}$ matrix, we can try to 
derive stochastic Hamiltonians following our previous works 
\cite{sy,akosy}. 
\item Extension to open strings: In particular, 
it is interesting to formulate some toy models for Dirichlet branes 
using the present method. It might be 
useful for studying the nonperturbative dynamics of the D-branes.
\item Inclusion of fermions: In particular, extension to 
supersymmetric models. 
\item Application to critical strings by extending the target 
space to higher dimensional spaces:  For this, however, we have to
deepen our understanding of the matrix-model formulation of
critical strings.  
\end{enumerate} 
In all of these possibilities, it would be more or less crucial to develop 
some powerful methods for 
treating the scaling limit in order to perform useful 
nonperturbative analysis on the basis of our models. 
We hope to return to some of the above issues in future works. 
  
\vspace{0.3cm}
The works of Y.O. and T.Y. are partially supported by 
the Japan Society for the Promotion of 
Science under the Predoctoral Research Program
 (No.08-4158) and the US-Japan Collaborative Program 
for Scientific Research, respectively.

\end{document}